\documentclass[twocolumn]{aastex63}

\received{June 1, 2021}
\revised{August 1, 2021}
\accepted{September 13, 2021}
\submitjournal{AJ}

\shorttitle{What's Behind the Elephant's Trunk?}
\shortauthors{Silverberg et al.}
\graphicspath{{./}{figures/}}
\begin{document}

\title{What's Behind the Elephant's Trunk? Identifying Young Stellar Objects on the Outskirts of IC 1396 \footnote{Based on observations obtained with XMM-Newton, an ESA science mission with instruments and contributions directly funded by ESA Member States and NASA.}}

\correspondingauthor{Steven M. Silverberg}
\email{ssilverb@mit.edu}

\author[0000-0002-3741-4181]{Steven M. Silverberg}
\affiliation{MIT Kavli Institute for Astrophysics and Space Research, 77 Massachusetts Avenue, Cambridge, MA 02139, USA}

\author[0000-0003-4243-2840]{Hans Moritz G{\"u}nther}
\affiliation{MIT Kavli Institute for Astrophysics and Space Research, 77 Massachusetts Avenue, Cambridge, MA 02139, USA}

\author[0000-0001-6072-9344]{Jinyoung Serena Kim}
\affiliation{Steward Observatory, Department of Astronomy, University of Arizona, 933 North Cherry Avenue, Tucson, AZ 85721, USA}

\author[0000-0002-7939-377X]{David A. Principe}
\affiliation{MIT Kavli Institute for Astrophysics and Space Research, 77 Massachusetts Avenue, Cambridge, MA 02139, USA}
%\collaboration{1}{(AAS Journals Data Scientists collaboration)}

\author[0000-0002-0826-9261]{Scott J. Wolk}
\affiliation{Smithsonian Astrophysical Observatory, MS 70, 60 Garden St., Cambridge, MA 02138}

%\author{Nicholas J. G. Cross}
%\affiliation{Wide-Field Astronomy Unit, Institute for Astronomy, School of Physics and Astronomy, University of Edinburgh, Royal Observatory, Blackford Hill, Edinburgh EH9 3HJ, UK}
%\nocollaboration{1}



\begin{abstract}
Empirically, the estimated lifetime of a typical protoplanetary disk is $<5-10$ Myr. However, the disk lifetimes required to produce a variety of observed exoplanetary systems may exceed this timescale. Some hypothesize that this inconsistency is due to estimating disk fractions at the cores of clusters, where radiation fields external to a star-disk system can photoevaporate the disk. To test this, we have observed a field on the western outskirts of the IC 1396 star-forming region with \textit{XMM-Newton} to identify new Class III YSO cluster members. Our X-ray sample is complete for YSOs down to $1.8\,M_{\odot}$. We use a subset of these X-ray sources that have near- and mid-infrared counterparts to determine the disk fraction for this field. We find that the fraction of X-ray-detected cluster members that host disks in the field we observe is $17_{-7}^{+10}\%$ (1$\sigma$), comparable with the $29_{-3}^{+4}\%$ found in an adjacent field centered on the cometary globule IC 1396A. We re-evaluate YSO identifications in the IC 1396A field using \textit{Gaia} parallaxes compared to previous color-cut-only identifications, finding that incorporating independent distance measurements provides key additional constraints. Given the existence of at least one massive star producing an external radiation field in the cluster core, the lack of statistically significant difference in disk fraction in each observed field suggests that disk lifetimes remain consistent as a function of distance from the cluster core.
\end{abstract}

%% Keywords should appear after the \end{abstract} cotestd. 


\section{Introduction} \label{sec:intro}
Nearby, young stellar clusters are the prime location for studies of the evolution of young stellar objects, specifically the evolution of their circumstellar disks. In the generally-accepted model of disk evolution \citep[e.g.][]{2011ARA&A..49...67W}, pre-main-sequence stars begin their lives surrounded by optically thick protoplanetary disks that accrete material onto their host stars. These objects are categorized as Class I and Class II young stellar objects (YSOs) based on the slopes of their infrared spectral energy distribution (SED) \citep{1987IAUS..115....1L}. Over time, these disks evolve and dissipate: photoevaporation, accretion, and the formation of gas giants work in concert to remove much of the gas from the disk, leaving behind a Class III YSO with an optically-thin debris disk \citep[e.g.][]{2018ARA&A..56..541H}, or no disk at all. 

Observational studies have shown that the disk fraction in clusters decreases with cluster age \citep[e.g.][]{2001ApJ...553L.153H,2009IAUS..253...37I,2009AJ....138.1116S}, and also correlates with stellar spectral type; later-type stars exhibit a higher frequency of disks than earlier-type stars in the same cluster \citep[e.g.][]{2006ApJ...651L..49C}. The timescales available for planet formation around young stars are constrained by these timescales for disk dissipation. While previous studies have shown that many primordial disks seemingly dissipate almost too quickly for the epoch of gas giant formation to occur, on the order of $\sim$3 Myr \citep[e.g.][]{2009IAUS..253...37I,2011ARA&A..49...67W}, more recent studies have shown that these timescales are slightly more relaxed from what was previously hypothesized, closer to 4-5 Myr for late-type stars \citep[e.g.][]{2016MNRAS.461..794P}. Despite this more relaxed timescale, current empirical disk dissipation timescales still do not entirely align with the prevalence of exoplanets observed in the galaxy \citep[e.g.][]{2013ApJ...766...81F,2020AJ....159..248K}.

\citet{2014ApJ...793L..34P} hypothesized that the difference in observed disk dissipation timescales and the prevalence of exoplanets is due to observational biases caused by the tendency of disk-frequency studies to focus on the center of clusters, where external radiation fields from massive stars in the cluster core will cause the disks to  dissipate more rapidly than if they were in a low-external-radiation environment \citep[e.g.][]{2013ApJ...774....9A}. \citet{2014ApJ...793L..34P} notes three biases in particular: (1) because small young clusters will disperse and mix with field stars after a few Myr, studies of clusters at ages 3-10 Myr focus on longer-lived massive clusters, which have a higher fraction of massive stars and thus a lower disk fraction; (2) radiation pressure pushes gas out of the cluster, and a large fraction of members become gravitationally unbound after 1-3 Myr, meaning that most of the remaining members come from the cluster core, where external radiation will dissipate disks more rapidly; (3) cluster members are more easily identified in the core of a cluster due to the apparent spatial over-density of stars. Additionally, the small field of view of modern telescopes makes obtaining a deep exposure of the full star-forming region infeasible.

These factors can combine to produce an underestimate of the true disk fraction in a cluster by focusing on that part of the cluster which intrinsically has a lower disk fraction than the rest of the cluster. To test the hypothesis that the tension in disk dispersal time vs the observed frequency of exoplanets is due to the selection bias of cluster studies, \citet{2014ApJ...793L..34P} suggest adopting a ``considerably larger field of view ($>20 \times 20$ pc) in older clusters would in principle reduce this effect.'' In contrast to this hypothesis, however, \citet{2014ApJ...787..109G} found that the disk fraction decreased as a function of distance from the cluster core in the Orion Nebula Cluster and NGC 2024, which would instead suggest that the outer regions of the cluster are more evolved than the core.

While disk-hosting cluster members are easily identified by their infrared excesses (due to the reprocessing of host-star light by the disk), determining disk fractions requires a robust method of identifying diskless members of young clusters, such as identification by X-ray emission which is saturated for young stars but decays with age \citep{2005ApJS..160..390P,2017MNRAS.471.1012B} such that field stars at the distance of the cluster would no longer be detected. This is commonly used to identify diskless cluster members \citep[e.g.][]{2008AJ....135..693W,2012AJ....144..101G}.

\begin{figure*}
    \centering
    \includegraphics[width=\textwidth]{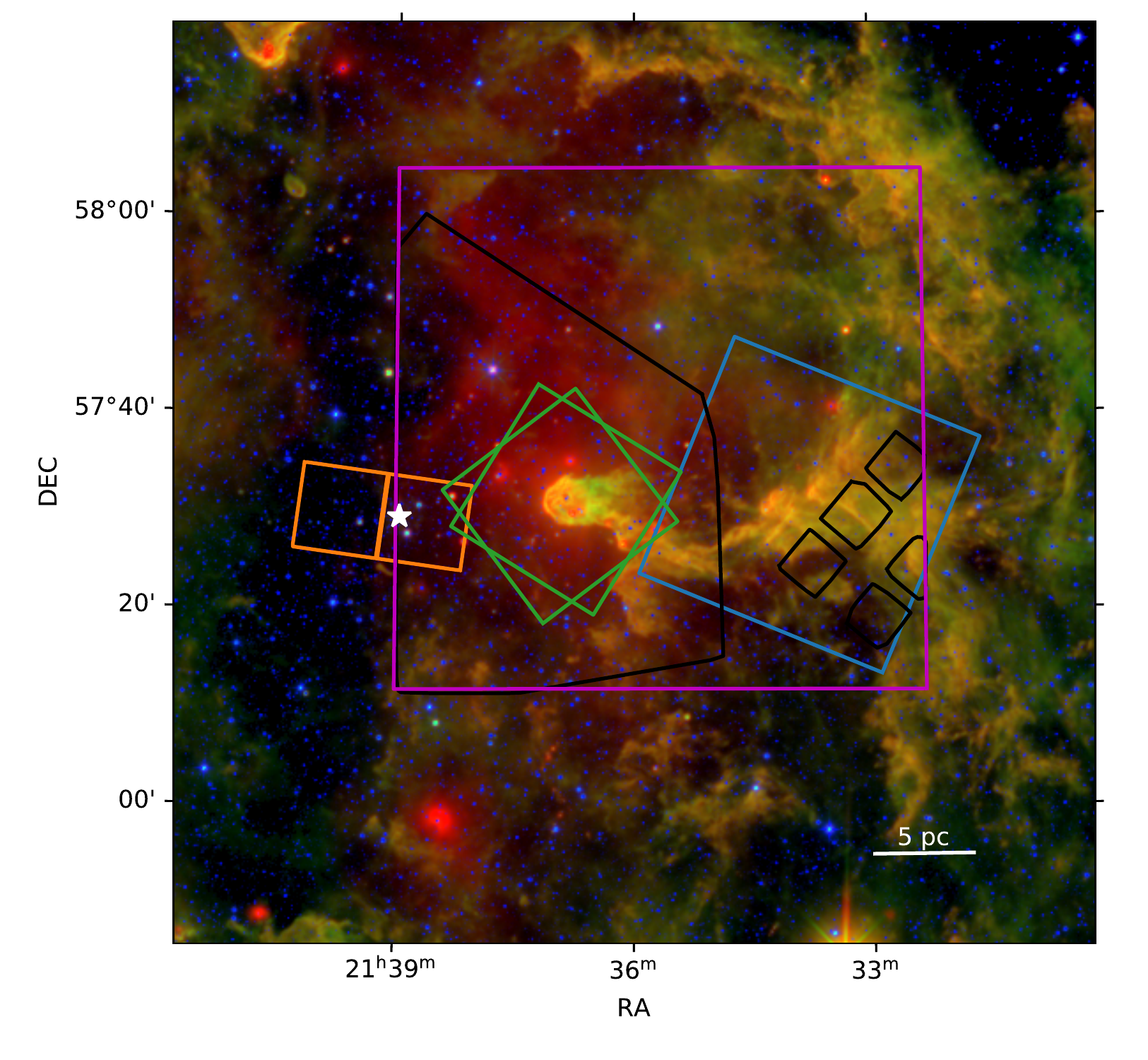}
    \caption{Selected X-ray and mid-IR pointings toward IC 1396, the Elephant Trunk nebula. Background is a false-color combination of the W1 (blue), W3 (green), and W4 (red) channels from the \textit{Wide-field Infrared Survey Explorer} (WISE). \textit{Chandra}/ACIS-I pointings toward the globule at the ``tip'' of the trunk \citep{2012MNRAS.426.2917G} are outlined in green. The orange outline shows the alignment of the inner two \textit{Chandra}/ACIS-S chips, analyzed in \citep{2009AJ....138....7M}. The blue outline shows the alignment of the \textit{XMM}-Newton PN camera during the new observations presented here. The MOS detectors cover a similar FOV. The blue points are point sources that are resolved in W1. The area covered by the UKIRT/WFCAM observations is outlined in magenta. Black outlines indicate the areas within the UKIRT/WFCAM field that are covered by at least one \textit{Spitzer}/IRAC band. The white star indicates the position of HD 206267, the spectroscopic O-star binary at the heart of this region.}
    \label{fig:Xraypointings}
\end{figure*}

The HII region IC 1396 is located at the edge of the Cepheus OB2 association, at a distance of 850 pc - 1 kpc \citep{2002AJ....124.1585C,2019A&A...622A.118S}. The central O star binary, HD 206267, ionizes the HII region, driving a shock front into the surrounding molecular cloud. HD 206267 is also the center of the 3.7-Myr open cluster Tr 37 \citep[and references therein]{2008hsf1.book..136K}, which is devoid of gas. HD 206267 has been the subject of several X-ray observations, which have been used to identify cluster members \citep{2009AJ....138....7M,2012MNRAS.426.2917G}. Three individual small globules in the region have also been observed \citep{2007ApJ...654..316G,2012MNRAS.426.2917G}, finding evidence of triggered star formation in the globules separate from the population of young stars associated with the open cluster Tr 37--for instance, the stellar population associated with the cometary globule IC-1396A is thought to be $\sim 1$ Myr old \citep{2012MNRAS.426.2917G} More than 800 members of the Tr 37/IC 1396 cluster have been identified through the use of X-ray observations \citep{2009AJ....138....7M,2012MNRAS.426.2917G}, spectroscopic surveys \citep{2002AJ....124.1585C,2006AJ....132.2135S,2013A&A...559A...3S}, infrared disk searches \citep{2004ApJS..154..385R,2006ApJ...638..897S,2009ApJ...702.1507M}, and H$\alpha$ photometry \citep{2011MNRAS.415..103B}.

In this paper, we present X-ray detections from an 87.3-ks exposure of the outer part of IC 1396, west of the core of the Trumpler 37 cluster and near the ``base'' of the Elephant's Trunk Nebula, with \textit{XMM-Newton}. For convenience, we refer to this field IC1396-West (or IC1396W) for the rest of the paper. We show the location of these observations relative to other X-ray observations of the region and the underlying nebular dust background in Figure \ref{fig:Xraypointings}. We use these data, in conjunction with archival optical, near-infrared, and astrometric data to identify likely members of the cluster. We identify striking patterns in the locations of young stars along the apparent edges of the dust clouds (as shown in WISE mid-infrared data). We also use this archival data in conjunction with previous observations of an adjacent region observed with \textit{Chandra}/ACIS-I centered on the cometary globule IC 1396A to compare our methods for disk identification to those of \citet{2012MNRAS.426.2917G}.

\section{Observations and Data Reduction} \label{sec:obs}

\subsection{XMM-Newton}

We obtained an 87.3 ks observation of the western outskirts of IC 1396 beginning on UT 2015 December 01 (\textit{XMM-Newton} observation 0762360101). This observation was designed to observe the outer part of the cluster behind the ``base'' of the Elephant's Trunk (the region highlighted in blue in Figure \ref{fig:Xraypointings}), specifically for purposes of identifying Class III sources. The observations used the medium thickness optical blocking filter. We used the pipeline reduction of these data produced as part of the full reprocessing of the \textit{XMM-Newton} archive for the 4XMM catalog in 2019\footnote{A custom configuration based on SAS version 18.0.0.}. Because there were no particularly bright targets in the center of the field of view, we do not incorporate data from either RGS camera. The reduction detects 152 X-ray sources.

\subsection{Near-Infrared Imaging from UKIRT}

A square degree of IC 1396 was monitored with the Wide-Field Camera (WFCAM) on the United Kingdom Infrared Telescope \citep{2007A&A...467..777C} over 21 epochs from July 18, 2014, to July 12, 2016, first presented in \citet{2019ApJ...878....7M}. While that work looked at variability between epochs, we use here the combined photometry across all 21 epochs, which allows us to obtain deeper photometry than that provided by the 2-Micron All-Sky Survey \citep[2MASS;][]{2006AJ....131.1163S}. Seeing ranged from $0.''6$ to $1.''4$, with average seeing of $0.''79$. We refer the reader to \citet{2019ApJ...878....7M} for details on the observations and reduction of this data. The pipeline-reduced data, which we use here, is available on the database WSERV9v20170222, at the WFCAM Science Archive\footnote{\url{http://http://wsa.roe.ac.uk/}} \citep{2008MNRAS.384..637H}.

\begin{figure}
    \centering
    \includegraphics[width=0.47\textwidth]{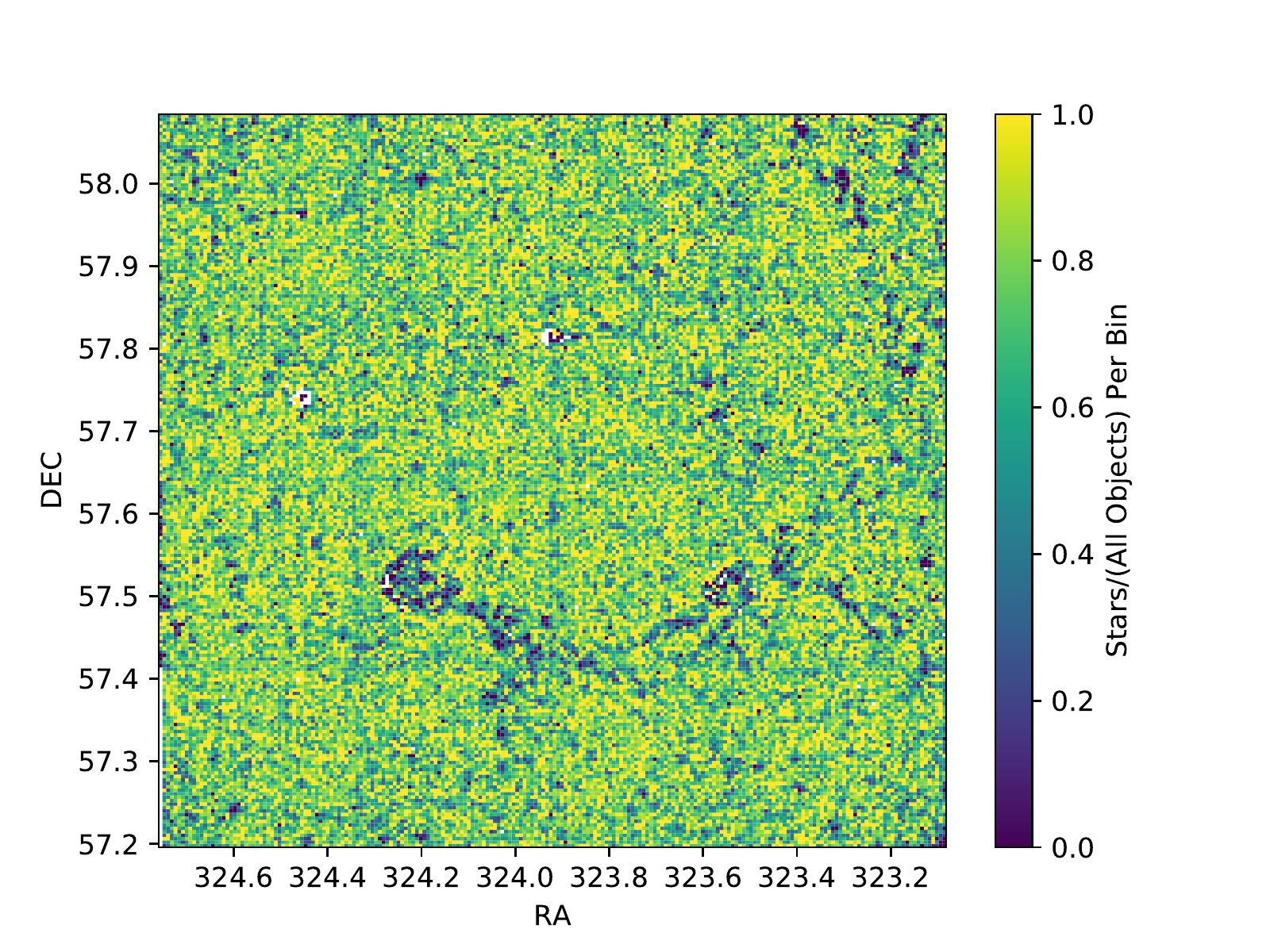}
    \caption{The ratio of objects classified as stars by the WFCAM pipeline to all objects shows that the fraction of objects labeled stars is generally constant with the exception of the Elephant Trunk nebula. This suggests that the difference is that the pipeline classifies objects differently in the region of the nebula. The faint grid pattern is due to source duplication from imperfect matching between observation sets.}
    \label{fig:ratio} 
\end{figure}

We noticed in our comparison of data that use of the star/galaxy probabilities provided by the WFCAM pipeline to eliminate sources without high probabilities of being stars \citep[as was done in][]{2019ApJ...878....7M} produced an under-density of sources identified as stars along the path of the dust bands observed in the WISE data, without a corresponding drop in the number of sources overall along the dust band but with a drop in number of total sources behind the cometary globule at the center of IC 1396-A. This most clearly appears when comparing the ratio of objects that are classified as stars in each spatial bin to the total number of objects in each bin, shown in Figure \ref{fig:ratio}. This suggests that the automated classification of star vs galaxy was affected by the dust (either due to foreground dust reddening the stars into colors more typically found in galaxies, or background dust producing a less defined point source). We therefore make no cuts based on the automated classification of each object in the UKIRT/WFCAM pipeline. We note the duplication of some sources in the overlap between regions observed at different times due to imperfect source matching; these observations typically had consistent (within uncertainties) photometry to each other at very low separations and came from different frame sets, suggesting that the sources were multiple detections of the same source. To eliminate these, we matched the UKIRT dataset to itself with a search radius of $0.''5$, below the typical spatial resolution of the observations, and for each set of duplicated sources selected one source identifier as the ``true'' source.

\subsection{Archival Data}

We retrieved photometry of the region observed in the UKIRT data from %Data Release 1 from the Panoramic Survey Telescope and Rapid Response System 1 \citep[PanSTARRS1 DR1;][]{2016arXiv161205560C}, 
the early third data release from ESA's \textit{Gaia} mission \citep[\textit{Gaia} EDR3;][]{2016A&A...595A...1G,2020arXiv201201533G}, the AllWISE catalog of data from the \textit{Wide-field Infrared Survey Explorer} (WISE) \citep{2013yCat.2328....0C}, and various pointed observations with the \textit{Spitzer Space Telescope} \citep{2004ApJS..154....1W}, obtained via the \textit{Spitzer} Heritage Archive. 

We use the UKIRT/WFCAM and 2MASS data as the backbone of our target matching. We searched the Spitzer Catalog for point sources matching the coordinates of our near-IR sources with a search radius of $1''$, chosen based on the typical $0''.8$ seeing of the UKIRT data and the \textit{Spitzer} point spread function, yielding 19,744 sources. We then cross-matched the near-IR data set with \textit{Gaia} EDR3, and AllWISE using the CDS xMatch functionality as implemented in \texttt{astroquery} \citep{2019AJ....157...98G}, again using a search radius of $1''$ for consistency with the \textit{Spitzer} crossmatch.

Finally, we used the CDS xMatch service to match the XMM and UKIRT/2MASS objects, with a match radius of $2''$ since the point spread function (PSF) of \textit{XMM-Newton} is considerably wider than the UKIRT and 2MASS PSFs. We examined by hand each of the matches between the \textit{XMM-Newton} and UKIRT sources to determine its relevance to our work (e.g.\ identifying the UKIRT source most likely to correspond to the X-ray data in cases where there were two or more UKIRT sources matched to the same X-ray source). The XMM field is aligned such that some of the field falls outside the region for which we have WFCAM data; we thus do not consider the six X-ray sources that fall outside of this range.

There are 47 X-ray sources in the \textit{XMM-Newton} observation without UKIRT/WFCAM counterparts. While most of these are likely extragalactic in origin, several are among the brightest X-ray sources, suggesting that they are  Galactic in origin. To check if these X-ray sources had near-IR coutnerparts too bright for UKIRT/WFCAM, we also cross-matched the \textit{XMM-Newton} data to 2MASS. However, we found that all XMM/2MASS matches in the UKIRT field also had UKIRT matches, suggesting that UKIRT saturation was not the cause of the lack of a near-IR counterpart. 

To compare our sample to other studies of this region, we also cross-matched X-ray sources from \textit{Chandra}/ACIS-I images of the cometary globule IC 1396A at the tip of the elephant's trunk \citep{2012MNRAS.426.2917G} to the UKIRT data, again using the CDS/Xmatch functionality. For convenience, we refer to this field as IC 1396A for the remainder of the paper. While this field does not cover the core of the Tr 37 cluster (which presumably formed out of the IC 1396 molecular cloud), it is more interior to the cluster than the IC 1396-West field. We do not examine X-ray data from the cluster core \citep{2009AJ....138....7M} because these \textit{Chandra} data have diminished sensitivity due to the presence of the \textit{Chandra} High-Energy Transmission Grating Spectrometer (HETG-S), and because the UKIRT field does not fully overlap with the cluster core. Considering just the IC 1396A field and the IC 1396-West fields provides near-complete coverage of the Elephant's Trunk from tip to where it meets the larger cloud, as depicted in Figure \ref{fig:Xraypointings}. We used the cross-matches to 2MASS identified by \citet{2012MNRAS.426.2917G} for those X-ray sources that did not have a UKIRT counterpart.

\subsection{Estimating the Cluster Distance with \textit{Gaia} EDR3} \label{sec:gaiaedr3}

We estimated the distance to the cluster using \textit{Gaia} EDR3 parallaxes to cluster members that have previously been identified via a variety of methods \citep{2002AJ....124.1585C,2005AJ....129..363S,2006AJ....132.2135S,2013A&A...559A...3S,2004ApJS..154..385R,2009ApJ...702.1507M,2011MNRAS.415..103B,2006ApJ...638..897S,2009AJ....138....7M,2012MNRAS.426.2917G}, following the methodology of \citet{2019A&A...622A.118S}. As with \citet{2019A&A...622A.118S}, we selected only those sources that had matching radii $<0.5$ arcsec, relative parallax uncertainties with a ratio of parallax uncertainty to parallax $\sigma_{\omega}/\omega < 0.1$, and proper motion uncertainties below 2 $\mathrm{mas\,yr^{-1}}$, to ensure that the \textit{Gaia} source was the same as the source in the existing catalog. We also required that for those objects which had both \textit{Gaia} DR2 and EDR3 parallax measurements, the estimated distance from EDR3 is consistent with its estimated distance from\textit{Gaia} DR2; this ensures that any differences in our estimate relative to \citet{2019A&A...622A.118S} are due to incorporating sources with new or improved-quality parallax measurements, rather than potential mismatches with \textit{Gaia} EDR3. We estimated distances following the prescription of \citet{2015PASP..127..994B}, using a characteristic length scale of $l = 1.35 \;\mathrm{kpc}$ in the prior based on the direction of observation \citep{2018AJ....156...58B}. 

We found an average distance to the cluster of $931_{-116}^{+159}$ pc, where the uncertainties indicate the 5th and 95th percentiles. This measurement is consistent with both the \citet{2019A&A...622A.118S} estimate of $945_{-73}^{+90}$ pc\footnote{We note that the value quoted here incorporates an additional prior to identify the highest-likelihood members, to enable a proper-motion study in \citet{2019A&A...622A.118S}. Removing this prior yields a distance $949_{-101}^{+161}$ pc (A. Sicilia-Aguilar, private communication), with uncertainties in line with those presented above.} and the previous value of 870 pc from \citet{2002AJ....124.1585C}. We note that our work here specifically does \textit{not} incorporate additional information beyond a literature identification of cluster membership and a parallax measurement; future work that incorporates more information from \textit{Gaia} (e.g. proper motion measurements) will likely produce a more precise result. However, this distance estimate will be sufficient for our purposes: excluding objects with distances that are inconsistent with cluster membership from our sample.

\section{Classifying YSOs}
\label{sec:classifications}

\begin{figure*}
    \centering
    \includegraphics[width=\textwidth]{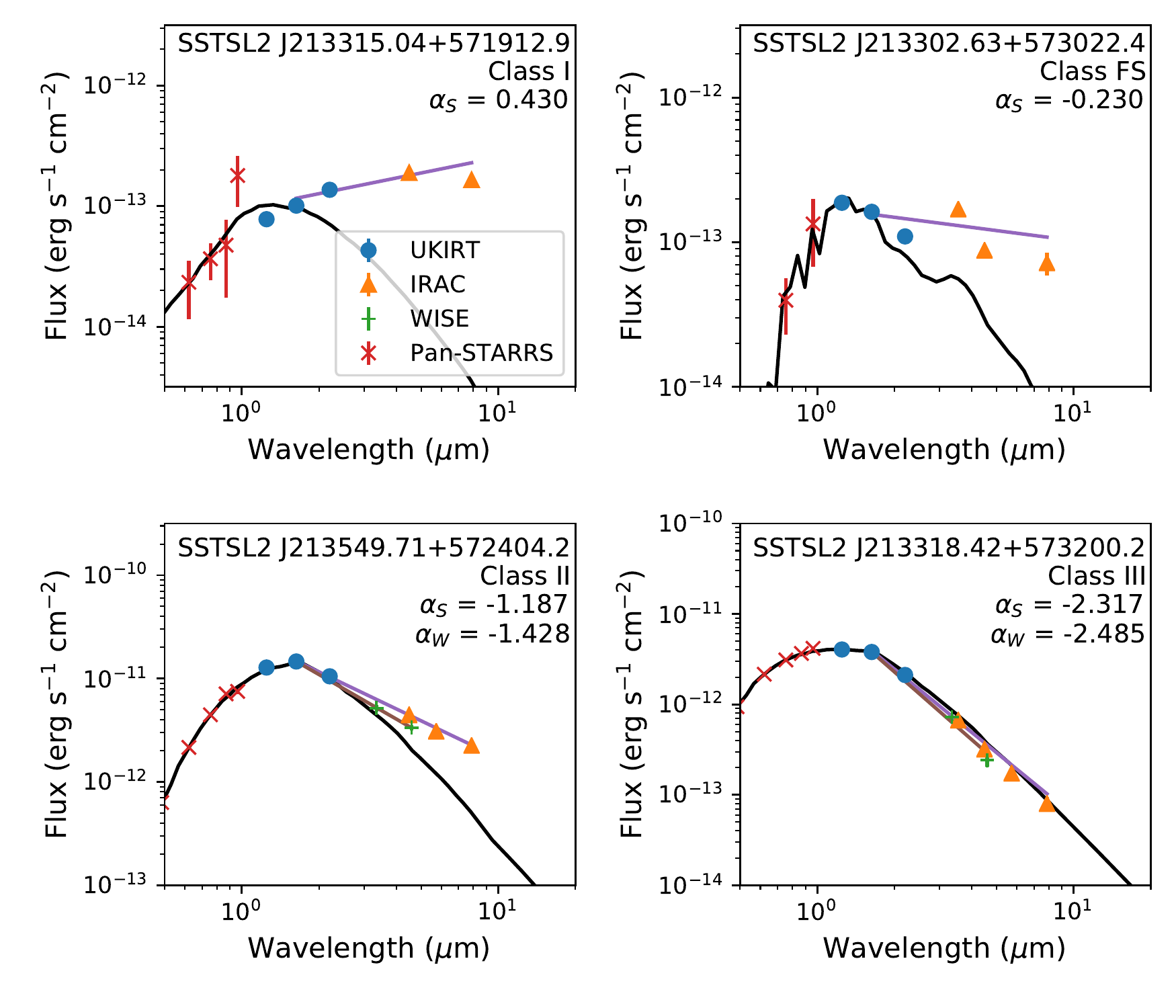}
    \caption{Example SEDs of each class of YSO. A photosphere-only model from \citet{2017A&A...600A..11R} is plotted in black for comparison. The best-fit power law to the \textit{Spitzer} data is shown in purple, while the best-fit power law to the AllWISE data (where available) is in brown. Error bars represent 3$\sigma$ uncertainties; these are typically too small to be seen in the UKIRT and \textit{Spitzer} data.}
    \label{fig:sed_examples}
\end{figure*}

Disks and circumstellar material are apparent as IR excess because the circumstellar material absorbs and re-radiates the stellar emission. The disk temperature decreases with distance from the star; thus the SED typically peaks at a few microns for disk sources, and at longer wavelengths for Class I sources, which are still embedded deep in their envelope. To assess the presence of a disk for each near-IR source with \textit{Spitzer} or AllWISE data, we determine the exponent $\alpha$ in a power-law fit to each object's spectral energy distribution (SED; e.g.\ Figure \ref{fig:sed_examples}), starting with the near-IR \textit{H} band from UKIRT/WFCAM, using maximum-likelihood estimation. We computed separate estimates for $\alpha$ based on  \textit{Spitzer}/IRAC data and AllWISE data, trading AllWISE's spatial coverage for the improved sensitivity of IRAC. Based on the full-frame images in AllWISE, we found that the 12- and 22-$\mu$m AllWISE data primarily detect the background emission from the nebula; we thus only used data from wavelengths shorter than 10 $\mu$m. We separate sources into YSO classes based on the boundaries on power-law fitting classifications outlined by \citet{2011ARA&A..49...67W}: Class I objects have power laws with exponents $\alpha > 0.3$; Class FS (flat-spectrum) objects have $0.3 > \alpha > -0.3$; Class II objects have $-0.3 > \alpha > -1.6$, and Class III objects have $\alpha < -1.6$. While there are some degeneracies between classes \citep[e.g. a Class II YSO viewed from high inclination can have a similar SED to a typical Class I YSO;][]{2006ApJS..167..256R}, this method provides a simple categorical assessment of the presence of circumstellar material without potentially over-fitting the data, as might happen with a more elaborate disk model. Example SEDs of each class are shown in Figure \ref{fig:sed_examples}. In the cases where an object had both \textit{Spitzer} and WISE data, we use the \textit{Spitzer}-derived value for $\alpha$, as WISE data is more likely to suffer from blends with nearby sources and extended cloud emission due to its larger PSF \citep[e.g.][]{2012MNRAS.426...91K,Kuchner2016,2018ApJ...868...43S}. We summarize our filtering procedure in Table \ref{table:filtering}.

\begin{deluxetable*}{lc}
\tablecaption{Filtering Procedure for Identifying YSOs  \label{table:filtering}}
\tablewidth{0pt}
\tablehead{\colhead{Categorization} & \colhead{Number of Sources}}
\startdata
UKIRT Sources with \textit{Spitzer} or \textit{WISE} Cross-matches & 6537 \\
Sources Outside the Cluster Distance per \textit{Gaia} & (5181) \\
\tableline
Remaining Sources & 1356 \\
\tableline
\tableline
\textit{Spitzer} Sources & 780 \\
UKIRT Sources Matched to Identical \textit{Spitzer} Sources\tablenotemark{a} & (18) \\
Remaining \textit{Spitzer} Sources & 762 \\
\hspace{5mm} \textit{Spitzer} Sources with $\alpha_{S} \geq -1.6$ & 213 \\
\hspace{10mm} Class I & 14 \\
\hspace{10mm} Class FS & 25 \\
\hspace{10mm} Class II & 174 \\
\hspace{5mm}\textit{Spitzer} Sources with $\alpha_{S} < -1.6$ & 549 \\
\tableline
\tableline
\textit{AllWISE}-Only Sources & 576 \\
UKIRT Sources Matched to Identical \textit{AllWISE} Source\tablenotemark{a} & (136) \\
\tableline
Remaining \textit{AllWISE} Sources & 440 \\
\hspace{5mm} \textit{AllWISE}-Only Sources with $\alpha_{W}\geq-1.6$ & 153 \\
\hspace{10mm} Class I & 72 \\
\hspace{10mm} Class FS & 26 \\
\hspace{10mm} Class II & 55 \\
\hspace{5mm}\textit{AllWISE} Sources with $\alpha_{W} < -1.6$ & 287 \\
\tableline
\tableline
Sources with $\alpha_{S,W} < -1.6$ & 836 \\
\hspace{5mm} Sources with $\alpha_{S,W} < -1.6$ at Cluster Distance & 345 \\
\hspace{5mm} Sources with $\alpha_{S,W} < -1.6$ Without Parallax Measurement & 491 \\
\tableline
\tableline
\tableline
Remaining Sources with \textit{XMM-Newton} Detections & 18 \\
\hspace{5mm} Disk Hosts Detected with \textit{XMM-Newton} & 3 \\
\hspace{10mm} \textit{Spitzer}-detected Disk Hosts with \textit{XMM-Newton} Detections & 1 \\
\hspace{10mm} \textit{AllWISE}-detected Disk Hosts with \textit{XMM-Newton} Detections & 2 \\
\hspace{5mm} Class III Candidates Detected with \textit{XMM-Newton} & 15 \\
\hspace{10mm} Class III Candidates at Cluster Distance with \textit{XMM-Newton} Detections & 13 \\
\hspace{10mm} Class III Candidates without Parallax Measurements with \textit{XMM-Newton} Detections & 2\\
\tableline
\tableline
\tableline
UKIRT Sources Without \textit{Spitzer} or \textit{WISE} Cross-matches Detected in X-rays\tablenotemark{b} & 40 \\
\enddata
\tablenotetext{a}{This refers to the case where more than one UKIRT source is matched to a given \textit{Spitzer} or \textit{AllWISE} source. In these cases, since we typically cannot determine whether one source is the ``correct'' match or if the mid-IR data is a blend of the two near-IR sources, we remove the UKIRT source from consideration.}
\tablenotetext{b}{This notes UKIRT sources without \textit{Spitzer} or \textit{WISE} cross-matches that were detected with \textit{XMM-Newton} and either \textit{Gaia} parallaxes that put them at the cluster distance, or no \textit{Gaia} parallax measurement.}
\end{deluxetable*}

Class III objects were initially identified as sources with SEDs that correspond to a stellar photosphere ($\alpha < -1.6$) that also emit X-rays (indicative of youth) and had a parallax per \textit{Gaia} EDR3 that placed them at the correct distance. Only thirteen Class III sources at the correct distance per \textit{Gaia} EDR3 were detected in the XMM data. Removing the \textit{Gaia} requirement while keeping the requirements on $\alpha$ and the presence of X-rays added an additional two sources.

In addition to assessing the slope of the objects' SEDs, we made additional cuts based on data from \textit{Gaia} EDR3, using the cluster distance estimate we derived in Section \ref{sec:gaiaedr3}. We initially selected only those objects with \textit{Gaia}-measured parallaxes with distances whose 90\% confidence intervals \citep[following][]{2015PASP..127..994B,2019A&A...622A.118S} overlap with the 90\% bounds identified in Section \ref{sec:gaiaedr3} for the cluster, to provide an independent constraint on what objects are or are not likely cluster members. We found that several sources along the boundary of the nebula do not have \textit{Gaia} parallaxes, which we expect is due to the combined effects of potential binarity, additional fuzziness of the images due to the background nebula, and incompleteness of \textit{Gaia} at these distances. We expect that many of these sources for which there is not a published parallax measurement from \textit{Gaia} will be included in future \textit{Gaia} releases. In the meantime, we assume that those objects that do not have \textit{Gaia} parallax measurements but \textit{do} have SED slopes indicative of a disk are associated with the nebula and the cluster, and thus include them in our measurements.

\begin{figure*}
    \centering
    \includegraphics[width=\textwidth]{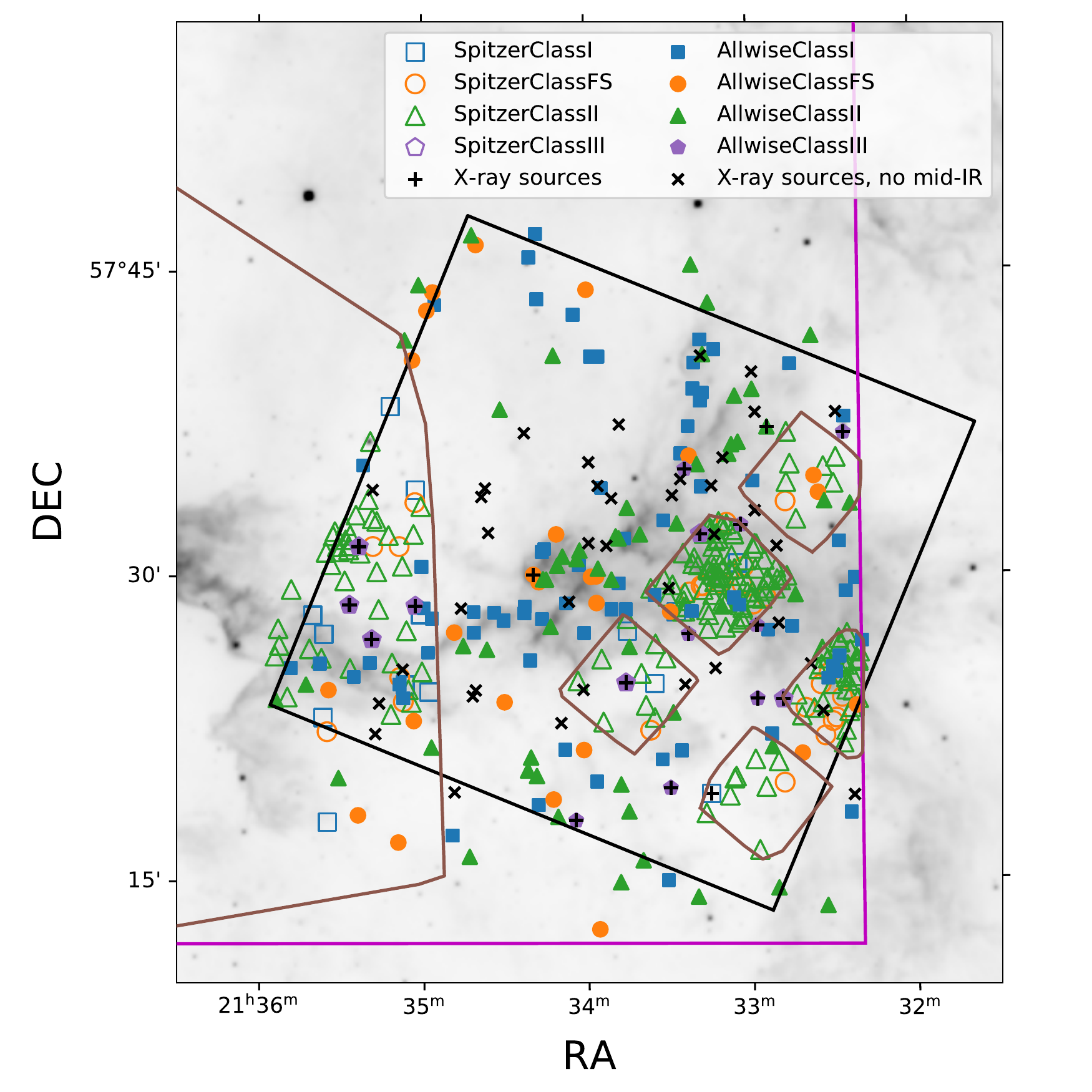}
    \caption{YSOs identified in the field of view of \textit{XMM-Newton} observations of the outskirts of IC 1396. Background is the 12-$\mu$m image of the region from \textit{WISE}. The field of the \textit{XMM-Newton} PN camera is outlined in black, while the UKIRT field is outlined in magenta. While more sensitive, \textit{Spitzer} observations are confined to specific pointings. Individual \textit{Spitzer}/IRAC pointings are indicated in brown. A subset of AllWISE-identified YSOs clearly fall along the nebula-wise dust lane concentrated in the center of the \textit{XMM-Newton} field. Class II is by far the most common classification. Interactive figure available in online version of journal; source classes can be interactively selected and deselected to improve visibility in dense regions, and sources are tagged individually with source ID, coordinates, $\alpha$, and X-ray flux.}
    \label{fig:YSOmap}
\end{figure*}

We map the locations of the YSOs identified in this sample in Figure \ref{fig:YSOmap}. We find that 5,707 UKIRT sources with matched Spitzer or WISE photometry in the IC 1396-West field have distances based on Gaia EDR3 parallaxes that are outside the cluster bounds derived in Section \ref{sec:gaiaedr3}, leaving 1356 UKIRT sources with WISE or Spitzer photometry to consider as potential cluster members. We will consider the sources with \textit{Spitzer} and WISE data separately in each of the following sections.

\subsection{Results with \textit{Spitzer}}

213 sources with \textit{Spitzer} photometry have $\alpha_{S} > -1.6$ from \textit{Spitzer}. 174 of these have $\alpha_{S}$ values that correspond to Class II YSOs, while 25 have slopes corresponding to Class FS and 14 have slopes corresponding to Class I. We list these in Table \ref{table:yso_candidates}. One of these sources is also detected with \textit{XMM-Newton}. %Only five of the sources detected in XMM are matched to UKIRT sources at the correct distance with associated sources in \textit{Spitzer}.

\begin{deluxetable*}{llll}
\tablecaption{YSOs in the \textit{XMM-Newton} Field of View  \label{table:yso_candidates}}
\tablewidth{0pt}
\tablehead{\colhead{Num} & \colhead{Label} & \colhead{Description} & \colhead{units}}
\startdata
1 & catUKIRT & UKIRT Source ID. Should be preceded by "44994077." & \nodata \\
2 & catSpitzer & \textit{Spitzer} source ID. & \nodata \\
3 & catWISE & \textit{WISE} source ID. & \nodata \\
4 & ra & Right Ascension (J2000) of UKIRT source. & degrees \\
5 & dec & Declination (J2000) of UKIRT source. & degrees \\
6 & plx & \textit{Gaia} EDR3 parallax. & mas \\
7 & plxerr & Uncertainty on \textit{Gaia} EDR3 parallax. & mas \\
8 & Jmag & UKIRT/WFCAM \textit{J} magnitude. & mags \\
9 & Jmagerr & Uncertainty of UKIRT/WFCAM \textit{J} magnitude. & mags \\
10 & Hmag & UKIRT/WFCAM \textit{H} magnitude. & mags\\
11 & Hmagerr & Uncertainty of UKIRT/WFCAM \textit{H} magnitude. & mags \\
12 & Kmag & UKIRT/WFCAM \textit{K} magnitude. & mags \\
13 & Kmagerr & Uncertainty of UKIRT/WFCAM \textit{K} magnitude. & mags \\
14 & I1flux & \textit{Spitzer}/IRAC \textit{I1} flux. & $\mu$Jy \\
15 & I1fluxerr & Uncertainty of \textit{Spitzer}/IRAC \textit{I1} flux  & $\mu$Jy \\
16 & I2flux & \textit{Spitzer}/IRAC \textit{I2} flux & $\mu$Jy \\
17 & I2fluxerr & Uncertainty of \textit{Spitzer}/IRAC \textit{I2} flux & $\mu$Jy \\
18 & I3flux & \textit{Spitzer}/IRAC \textit{I3} flux & $\mu$Jy \\
19 & I3fluxerr & Uncertainty of \textit{Spitzer}/IRAC \textit{I3} flux & $\mu$Jy \\
20 & I4flux & \textit{Spitzer}/IRAC \textit{I4} flux & $\mu$Jy \\
21 & I4fluxerr & Uncertainty of \textit{Spitzer}/IRAC \textit{I4} flux & $\mu$Jy \\
22 & W1mag & \textit{WISE} \textit{W1} magnitude. & mags \\
23 & W1magerr & Uncertainty of \textit{WISE} \textit{W1} magnitude. & mags \\
24 & W2mag & \textit{WISE} \textit{W2} magnitude. & mags \\
25 & W2magerr & Uncertainty of \textit{WISE} \textit{W2} magnitude. & mags \\
26 & alphaS & \textit{Spitzer}-derived SED slope $\alpha_{S}$ & \nodata \\
27 & alphaW & \textit{WISE}-derived SED slope $\alpha_{W}$ & \nodata \\
28 & class & Adopted YSO class & \nodata \\
29 & xrayflux & Combined (EPIC) X-ray flux (0.2-12 keV) & erg s$^{-1}$ cm$^{-2}$ \\
30 & xrayfluxerr & Uncertainty on combined X-ray flux & erg s$^{-1}$ cm$^{-2}$ \\
31 & HR & Hardness ratio. & \nodata \\
32 & HRerr & Uncertainty on hardness ratio. & \nodata \\
33 & PrevId & Previous references to this source. & 1,2,3\tablenotemark{1} \\
\enddata
\tablecomments{Machine-readable versions of the full table are available. Here, we list the columns and descriptions of the data in that table.}
\tablenotetext{1}{References: (1) \citet{2012MNRAS.426.2917G}; (2) \citet{2011MNRAS.415..103B}; (3) \citet{2005AJ....130..188S}}
\end{deluxetable*}
%Of the 4,989 UKIRT sources with associated AllWISE sources in the XMM field, only 618 do not have a parallax measurement that places their distance outside the expected bounds of the cluster. Of these 618, 359 have SEDs with power-law slopes of $\alpha > -1.6$. 124 of these are Class II YSOs, while 171 are Class I and 64 are Class FS.

%The primary method of detecting Class III objects is identifying sources with SEDs that correspond to a stellar photosphere ($\alpha < -1.6$) that also emit X-rays (indicative of youth). Only nine sources at the correct distance per \textit{Gaia} DR2 were detected in the XMM data.

345 sources with mid-IR data from \textit{Spitzer} or \textit{WISE} have SED slopes corresponding to a bare photosphere ($\alpha <= -1.6$) and \textit{Gaia}-parallax-based distances that place them at the correct distance for cluster membership, while an additional 491 have SED slopes corresponding to a bare photosphere and no parallax measurement. Of these, fifteen also have X-ray detections with \textit{XMM-Newton}, making them Class III YSO candidates. We list these in Table \ref{table:yso_candidates}. The others require additional observational data, such as deeper X-ray observations or the detection of lithium in their spectra, to determine their youth.

\subsection{Assessing Confusion and Contamination with AllWISE} \label{sec:AllWISE}

288 sources have $\alpha > -1.6$ from AllWISE. We checked each of these sources that did not also have \textit{Spitzer} data to determine whether the source was likely to be blended, contaminated, or confused, based on whether it was matched to multiple UKIRT sources using the $1''$ search radius. We excluded any UKIRT sources for which the same AllWISE source was matched to multiple UKIRT sources. This leaves 153 sources with $\alpha_{W} > -1.6$ that we believe to be uncontaminated. Of these, 72 have SED slopes corresponding to Class I; 26 have SED slopes corresponding to Class FS; and 55 have SED slopes corresponding to Class II. Two of these sources (one Class FS YSO, and one Class II YSO) are also detected with \textit{XMM-Newton}. These are listed in Table \ref{table:yso_candidates}.

We expect that there will be some differences between detections with \textit{WISE} and \textit{Spitzer}, due to differences in the \textit{WISE} and \textit{Spitzer} PSFs, observation strategies (\textit{Spitzer} provides deeper observations at the expense of spatial coverage), and data reduction (e.g. background subtraction).

To check whether we could reasonably combine estimates of $\alpha$ from \textit{Spitzer} and \textit{AllWISE}, we considered the 105 sources in IC1396-West that are detected with both telescopes. Of these, 99 have $\alpha$ in each telescope that correspond to the same YSO class. Two Class II sources in \textit{Spitzer} have Class III slopes in \textit{WISE}, and one system has a Class II slope in \textit{Spitzer} but a Class FS slope in \textit{AllWISE}. Most notably, three sources with $\alpha_{S}$ corresponding to Class III have $\alpha_{W}$ corresponding to disk-hosting classes---i.e. three Class III sources are mis-classified as disk hosts using WISE data.  
%Clarify that this is not based on central disk clearing etc.; rather it's differences in the Spitzer and AllWISE data at the same wavelengths (contamination, background subtraction, etc.)
%We only have complete sets of WISE data. Where we have both, there's 95% agreement.
If we assume similar statistics hold for the rest of the field, $95.2^{+1.7}_{-2.5}\%$ of YSO classifications in the AllWISE-only sample would be classified the same way by \textit{Spitzer}, indicating that we can reliably compare results from each telescope despite their differences

\subsection{X-ray detected cluster members with mid-IR data}
Fifteen UKIRT/WFCAM sources with mid-IR matches and $\alpha < -1.6$ are detected with \textit{XMM-Newton}, as are three UKIRT/WFCAM sources with mid-IR matches and $\alpha > -1.6$. These are listed in Table \ref{table:yso_candidates}. This yields a disk fraction for X-ray-detected cluster members of $17_{-7}^{+10}\%$, comparable within uncertainties to both the $29_{-3}^{+4}\%$ claimed by \citet{2012MNRAS.426.2917G} for the IC 1396A field between IC 1396-West and the core, and to the $20_{-7}^{+10}\%$ disk fraction for X-ray-detected cluster members in the core of the Tr 37 cluster by \citet{2009AJ....138....7M}. The uncertainties on these disk fractions was determined using the Wilson score interval for uncertainty in binomial proportions \citep{Wilson1927} with $z=1$. In summary, between disk-hosting sources that we detect in the archival data, and the Class III sources detected with \textit{XMM-Newton}, we identify 381 new YSOs in IC 1396 with mid-IR data.

\section{Discussion} \label{sec:discussion}

\subsection{Completeness of Our Sample}
\label{sec:completeness}
Many factors limit our ability to effectively identify \textit{all} cluster members in the region where IC 1396-West overlaps with the UKIRT/WFCAM data. Higher-mass stars are intrinsically brighter than their lower-mass counterparts, making them more detectable in X-rays. This is compounded by the suppression of X-ray emission due to accretion in disk hosts, as higher-mass stars dissipate their disks more quickly than their lower-mass counterparts \citep[e.g.][]{2006ApJ...651L..49C}. An X-ray-selected sample is thus biased towards higher-mass stars which are more likely to be diskless, which will result in a disk fraction lower than the true disk fraction for the region.

Simply identifying additional disk hosts provides another observational bias. The SED of a disk host is fairly conspicuous even in faint mid-IR data, such that we can detect a disk around a low-mass star just as readily as around a high-mass star. Similarly, a disk indicates youth just as readily as X-ray emission, so we can readily identify disk-hosting low mass stars without X-ray emission. However, only including sources with either an X-ray detection or a disk biases our search heavily in favor of finding a disk fraction larger than the true disk fraction, as this would include the low-mass disk hosts while excluding the low mass non-disk-hosts.

Additionally, the effects of the surrounding nebula could lead to misidentification of an embedded cluster member as a background extragalactic source. We can illustrate this by considering the case of a hypothetical single $1.0 M_{\odot}$ YSO at 1000 pc (within the expected range for cluster distance, but further toward the back of the cluster), at varying levels of extinction (and thus varying degrees of interstellar absorption). We use the colors of a $1.0 M_{\odot}$ pre-main sequence star from \citet{2013ApJS..208....9P}, and assume an unabsorbed flux for a $1.0 M_{\odot}$ weak-lined T Tauri star from the $L_{X} - M$ relationship of \citet{2007A&A...468..425T}, following an APEC model with $\log(T) = 7.05$. We assume for illustrative purposes the relationship between column density $N_{H}$ and $A_{V}$ from \citet{2009MNRAS.400.2050G}: $N_{H} = (2.21 (\pm 0.09) \times 10^{21}) A_{V}$. In Table \ref{table:extinction}, we present five cases for this source: no foreground material, the typical cluster extinction per \citet{2005AJ....129..363S}, and $A_{V} = $ 5, 10, and 20, corresponding to varying levels to which the source is still embedded in its natal envelope. As these demonstrate, a YSO could be faint enough to go undetected in $V$-band observations, and on the borderline of detection in UKIRT/WFCAM $K$-band, while still producing detectable X-rays. This could account for some portion of detected X-ray sources without a near-IR counterpart.

\begin{deluxetable}{ccccccc}
\tablecaption{Predicted Characteristics of a $1.0 M_{\odot}$ YSO at Various Extinction Levels \label{table:extinction}}
\tablehead{\colhead{} & \colhead{$N_{H}$} & \colhead{Count Rate} & \multicolumn{4}{c}{Magnitudes} \\
\colhead{$A_{V}$} & \colhead{($10^{21} cm^{-2}$)} & \colhead{(cts/s)} & \colhead{$V$} & \colhead{$K$} & \colhead{$I2$} & \colhead{$W2$}}
\startdata
0    &  0    & $3.0\times10^{-2}$ & 15.7 & 12.8 & 12.77 & 12.76 \\
1.56 &  3.5 & $1.1\times10^{-2}$ & 17.3 & 13.0 & 12.85 & 12.84 \\
5    & 11.1 & $3.8\times10^{-3}$ & 20.7 & 13.4 & 13.03 & 13.01 \\
10   & 22.1  & $1.4\times10^{-3}$ & 25.7 & 14.0 & 13.30 & 13.27 \\
20   & 44.2  & $5.2\times10^{-4}$ & 35.7 & 15.2 & 13.82 & 13.78 \\
\enddata
\end{deluxetable}

More subtly, relying on \textit{Spitzer} data leaves us only with data in the specific pointings available. While it has more spatial coverage, AllWISE is also magnitude limited. Heavily-extincted sources may thus go undetected in IR, while their X-rays are detected as hard sources without an IR counterpart and interpreted as a background AGN. The wide PSF of AllWISE \citep[$\sim 6''$ at W2;][]{2010AJ....140.1868W} also means that inevitably several potential good disk candidates are removed from consideration due to confusion and contamination (see Section \ref{sec:AllWISE}). 

Astrometry, independent of X-ray and IR emission, could potentially be used to identify cluster members---ideally, we would be able to map the bulk motion of known cluster members, and then algorithmically find objects with similar motions. However, \textit{Gaia} EDR3 is only complete to a magnitude limit $G = 17$ \citep{2020arXiv201201533G}, with a poorly-defined faint limit beyond that; if we conservatively assume incompleteness below $G = 17$, an extinction $A_{G} \simeq 2$, and a distance to the cluster of 931 pc, \textit{Gaia} data will be incomplete for cluster members with $M_{G} \gtrsim 5.2$. This would bias the sample both toward targets at the front of the cluster (i.e.\ at distances $< 931$ pc) due to their higher apparent brightness and minimized extinction from the nebula, and toward intrinsically brighter (i.e. high mass) sources.

While the issues all combine to provide an incomplete picture of the cluster in this region, it is worth noting that past surveys should suffer from similar issues for similar reasons. We thus conclude that our search with X-rays is broadly comparable to past X-ray-based searches for cluster members in IC 1396. For the purposes of estimating the disk fraction of cluster members, we focus on the X-ray-detected cluster members, as this is the surest way to ensure that we are comparing similar sub-populations of the cluster. We acknowledge that this disk fraction estimate is at best a lower limit on the true disk fraction of the cluster, given the number of low-mass stars we do not consider here; however, it is the best estimate available with the data we have in-hand. While we do not incorporate the disk-hosting YSOs we identify that are not detected in X-rays into our disk fraction estimate, we present their identification here so that a more accurate disk fraction can be computed as more complete data sets become available.

\subsection{Remaining X-ray Sources}
\label{sec:hardness_ratios}

Only 18 of the 152 X-ray sources in IC 1396-West are identified clearly as disk-hosting or non-disk-hosting cluster members. While six of the remaining 134 are not included because they do not overlap with the UKIRT/WFCAM observations, it behooves us to consider the remaining 128 sources.

47 XMM sources in the UKIRT field have neither a UKIRT nor 2MASS counterpart within $2''$. We expect that these are likely either extragalactic sources, or cluster sources embedded deep enough within the nebula that they are undetected in UKIRT due to extinction. For the full \textit{XMM-Newton}/EPIC field of view, we would expect $\sim 78$ extragalactic background sources with fluxes $>2.5\times10^{-14} \mathrm{erg\,s^{-1}\,cm^{-2}}$ \citep{2003ApJ...588..696M}; our number of 47 unmatched sources in the UKIRT/XMM overlap region (more than half of the full field of view) is consistent with that number. We do not detect as many sources that can be interpreted as extragalactic background sources as were found in the IC 1396A field \citep{2012MNRAS.426.2917G}; we expect that this is due to the apparent higher column density of the nebula in IC 1396-West compared to the IC 1396A field (Figure \ref{fig:Xraypointings}), the higher background level for \textit{XMM-Newton} compared to \textit{Chandra}, the larger PSF for \textit{XMM-Newton} resulting in faint background sources going unresolved, and the lower sensitivity threshold we find for the \textit{XMM-Newton} observations compared to the \textit{Chandra} observations (Section \ref{sec:XLFcompleteness}).

55 XMM sources have UKIRT counterparts that either do not have matches to \textit{Spitzer} or AllWISE sources, or do not have a valid slope solution. Of these, 16 have \textit{Gaia} parallax measurements that put them outside the cluster, while two have parallaxes that put them within the cluster and 38 have no measured parallaxes. These 40 sources (two with in-cluster parallaxes, 38 without measured parallaxes) are likely cluster members, bringing the total number of YSOs we identify to 421, but we lack sufficient data to determine if they host disks or not.

23 XMM sources have UKIRT counterparts and matches to \textit{Spitzer} or AllWISE, but have \textit{Gaia} parallaxes that put them outside the cluster. These are a mix of 12 foreground sources, two background sources, and nine sources where the range between the 5th and 95th percentile values of the distance distribution exceeds 1000 pc. The nine sources with highly uncertain distance estimates are almost certainly background sources; however, in some cases their highly uncertain distance estimates overlap with the distance estimate for the cluster, so they are treated separately.

This leaves twenty sources with UKIRT matches, \textit{Spitzer} or AllWISE data matched to the UKIRT source, and a parallax that does not exclude them from the cluster. In two of these cases, the AllWISE match is matched to more than one UKIRT source, and these are thus excluded (as in Section \ref{sec:AllWISE}), to produce our final sample of 18.

It is worth noting that the distributions of the near-IR ($K$-band) brightness of the X-ray-detected cluster members with and without mid-IR data are distinct. The sources with mid-IR data are in general much brighter---the median UKIRT/WFCAM $K$ magnitude of the sources with mid-IR data is 12.539, while the median $K$ magnitude for the sources without mid-IR data is 18.304. This is not overly surprising, as one would expect brighter sources to be those detected with WISE in the regions where \textit{Spitzer} data is unavailable. It does, however, indicate that the sources without mid-IR data are likely of lower mass that the sources with disks, likely are at the far side of the cluster rather than the near side, and thus likely are more embedded in the nebula as well.

\begin{figure}
    \centering
    \includegraphics[width=0.47\textwidth]{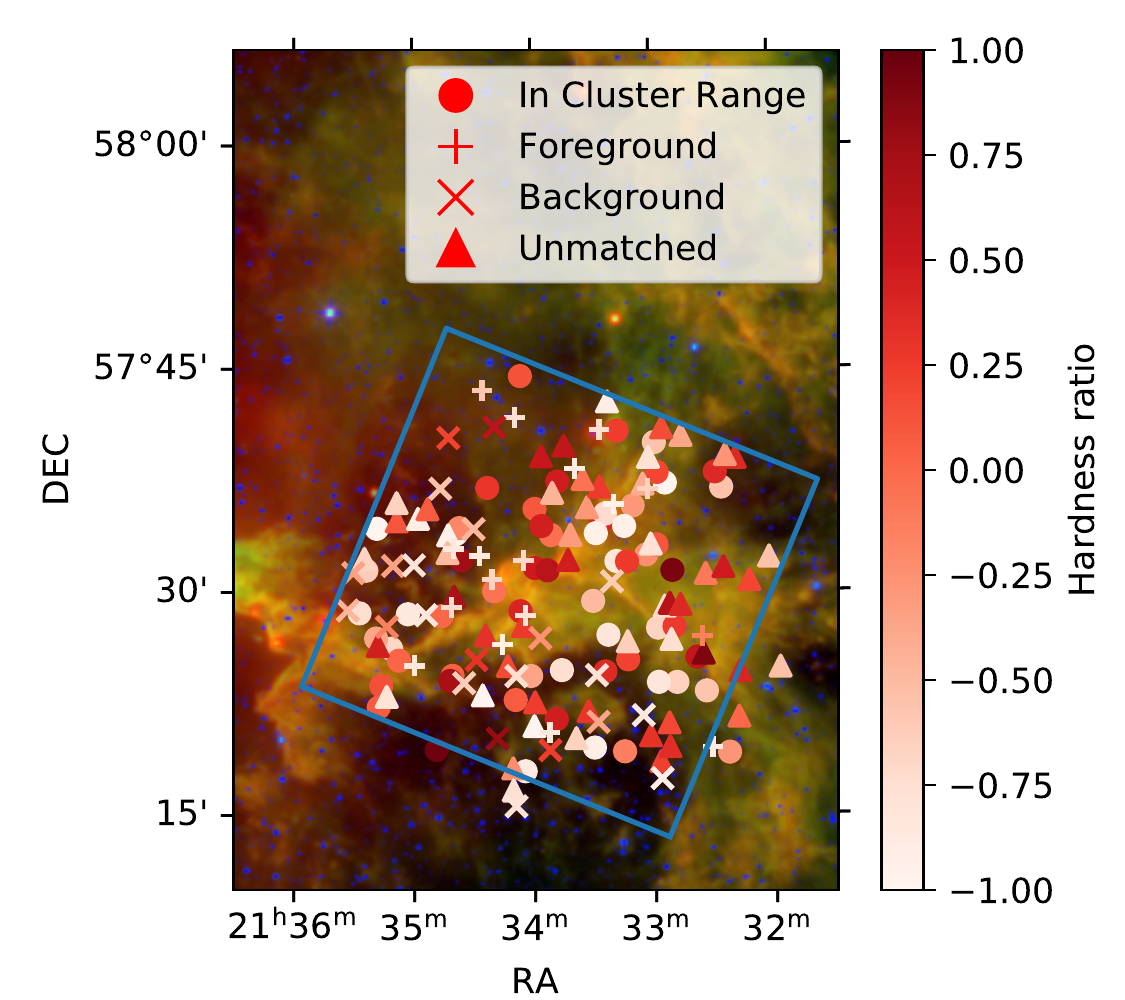}
    \caption{Hardness ratios as measured by \textit{XMM-Newton}. Color gradient indicates the hardness ratio from white (soft) to red (hard). Different markers indicate objects in different categories based on distance derived from \textit{Gaia} EDR3: in range for cluster membership, foreground, background, and sources for which \textit{Gaia} does not provide useful information (sources that are undetected, do not have a measured parallax, or which have a distance with uncertainty greater than 1000 pc.}
    \label{fig:hardness_ratios}
\end{figure}

To evaluate these X-ray sources independently of their match to UKIRT, we computed hardness ratios for each source, using the broad bands defined in the \textit{XMM-Newton} pipeline (band 6 $=$ 0.2-2 keV; band 7 $=$ 2-12 keV):

\begin{equation}
    HR = \frac{B_{H} - B_{S}}{B_{H}+B_{S}},
\end{equation}

\noindent
where $B_{H,S}$ are the count rates in the hard and soft bands, respectively. These are shown in comparison to the WISE image in Figure \ref{fig:hardness_ratios}.

As expected, foreground sources are the softest sources. The background sources are not exclusively hard, though this varies positionally, with softer background sources appearing at areas of minimal dust emission, as expected for Galactic background sources. The cluster members show a wide variety of hardnesses; some are among the hardest sources in the field. These appear to be spatially coincident with the star-forming nebular cloud; harder cluster members typically fall along the dust band, indicating a higher degree of embeddedness. The unmatched sources vary in hardness as well, with some soft sources appearing in regions of low nebular emission, as well as the harder emission expected of extragalactic sources. The appearance of a set of soft unmatched sources along the nebula as it branches to the northwest (starting at approximately right ascension 21h33m42s and declination $57^{\circ}32'06''$) suggests a potential additional population of cluster members that are not detected in the infrared data, either due to extinction from the nebula or intrinsic faintness. 

%As expected, for those objects with useful parallax information, hardness ratios increase as a function of distance---harder sources are generally more distant than soft sources. As expected, the sources not on the nebular cloud (e.g. those at the upper corner of the \textit{XMM-Newton} observation) are harder and more distant, suggesting they form the extragalactic and distant galactic background.

%However, \textit{Gaia} data has limitations at these distances. Several sources with limited or no information from \textit{Gaia} have hardness ratios comparable to our observed cluster members, and appear to fall on the nebular cloud, suggesting that they might be cluster members as well. Additional data will be necessary to absolutely confirm these sources as members. These X-ray detections that do not match to \textit{Gaia} data establish the utility of these observations going forward. X-ray observations of these systems mark them as possible YSO candidates, where reliance solely on \textit{Gaia} data would lead to them not being identified. An approach that combines these aspects shows the necessity of both pieces.

\subsection{Distribution of Cluster Members Through the Nebula}
\label{sec:distribution}

We find that only $17_{-7}^{+10}\%$ of likely cluster members in the IC 1396-West field with X-ray detections and available mid-IR data have disks. While this is not statistically different from the $29_{-3}^{+4}$\% found by \citet{2012MNRAS.426.2917G} or the $20_{-7}^{+10}\%$ found by \citet{2009AJ....138....7M}, it qualitatively suggests that the disk frequency \textit{diminishes} as a function of distance from the cluster core, as found for the ONC by \citet{2014ApJ...787..109G}, though some of this difference is likely due to the difference in X-ray sensitivity (Sections \ref{sec:completeness} and \ref{sec:XLFcompleteness}). While our statistics are not robust enough to provide a quantitative analysis of how the disks are distributed, we can qualitatively evaluate the distribution of disks through IC 1396-West, as depicted in Figure \ref{fig:YSOmap}, keeping in mind the selection effects of the observed distribution due to the targeted nature of the \textit{Spitzer} observations. We can also qualitatively assess the distribution of X-ray-identified likely cluster members through the nebula in conjunction.

We clearly see that disks tend to fall along the dust bands in the AllWISE data, which is borne out in the central \textit{Spitzer} field (which had coverage in both I1 and I2)---more sources are identified in the central pointing of the \textit{Spitzer} data, which falls on the nebula, and they generally tend to be disk sources. We note that the higher density of sources detected by \textit{Spitzer} is due to its deeper observations---the faintest \textit{Spitzer} sources in this field are an order of magnitude fainter in observed brightness than the faintest AllWISE sources.

There is a void of disk hosts and non-disk sources in the northern portion of the \textit{XMM-Newton} field, save for a small group of disk sources in the upper corner that are not detected in X-rays, despite seemingly low nebular emission compared to the main ``trunk.'' As fewer disks are expected in regions of lower nebular emission, the presence of these sources is intriguing. The group of disks seems potentially connected to a cluster of three X-ray sources without IR data (one in the cluster, two others unmatched to UKIRT), suggesting a filament of ongoing formation.

Surprisingly, many of the AllWISE-detected disk sources along the trunk are classified as Class I, the youngest type considered here. This suggests potential ongoing new star formation, extending the ``triggered star formation'' identified by \citet{2012MNRAS.426.2917G} to a further distance from the cluster core. This supports the hypothesis of \citet{2014ApJ...793L..34P}---it will take longer for Class I YSO disks to evolve into Class II YSOs and then eventually dissipate their disks than it will take Class II YSO disks to dissipate, so the Class I disks identified here should remain present (albeit evolving) for an extended period of time.

There is also an intriguing band of X-ray-detected cluster members tracing the gap between the southern edge of the main trunk of the nebula and a smaller dust cloud. This range includes an X-ray-detected Class III source in a \textit{Spitzer} field. The relative over-abundance of X-ray-detected sources in this belt compared to the dust band above, combined with the higher concentration of disk sources along the belt, suggests that these sources are likely older, having already formed, evolved, and dissipated their disks. Similarly, the disks identified in this region are primarily \textit{Spitzer}-detected Class II YSOs, suggesting a more-evolved state of evolution than the Class I sources found along the dust band.

\citet{2014ApJ...793L..34P} hypothesized that disk fractions would be higher outside of the core of a massive cluster than in the cluster core, due in part to the enhanced contributions of external photoevaporation on the dissipation of the disk. However, Tr 37 and IC 1396 have an O-star binary at the core (HD 206267), which should contribute both to disk photoevaporation and dissipation of the nebula. We can see the nebular dissipation effects in the ``bubble'' appearance of the background gas and dust (Figure \ref{fig:Xraypointings}). However, the disk fraction we find here at minimum indicates no increase in the disk fraction further away from the cluster core. While our data are not robust enough to completely disprove this hypothesis, they do suggest that if there is a disk dissipation gradient, it is in the opposite direction of what \citet{2014ApJ...793L..34P} hypothesize.

%Notably, many of the Class III sources that appear in the \textit{Spitzer} fields are also detected as Class III sources by AllWISE, while this is not necessarily true for disk-hosting sources. This could be due to the differences in how we identified these two types of source: we require Class III sources to have a measured parallax \textit{and} an X-ray detection, which might bias this selection to both intrinsically brighter sources and sources that are less deeply embedded in the nebula. Both conditions would make these targets easier to detect by \textit{Gaia} and \textit{XMM-Newton}, and this difference in brightness would make these sources detectable by AllWISE. As the quality of \textit{Gaia} measurements increases with future data releases, we would expect more faint sources to be detected at this distance, improving the completeness of the sample and potentially enabling further detection of Class III YSOs.

\subsection{Completeness of the X-ray Luminosity Function}
\label{sec:XLFcompleteness}
\begin{figure}
    \centering
    \includegraphics[width=0.47\textwidth]{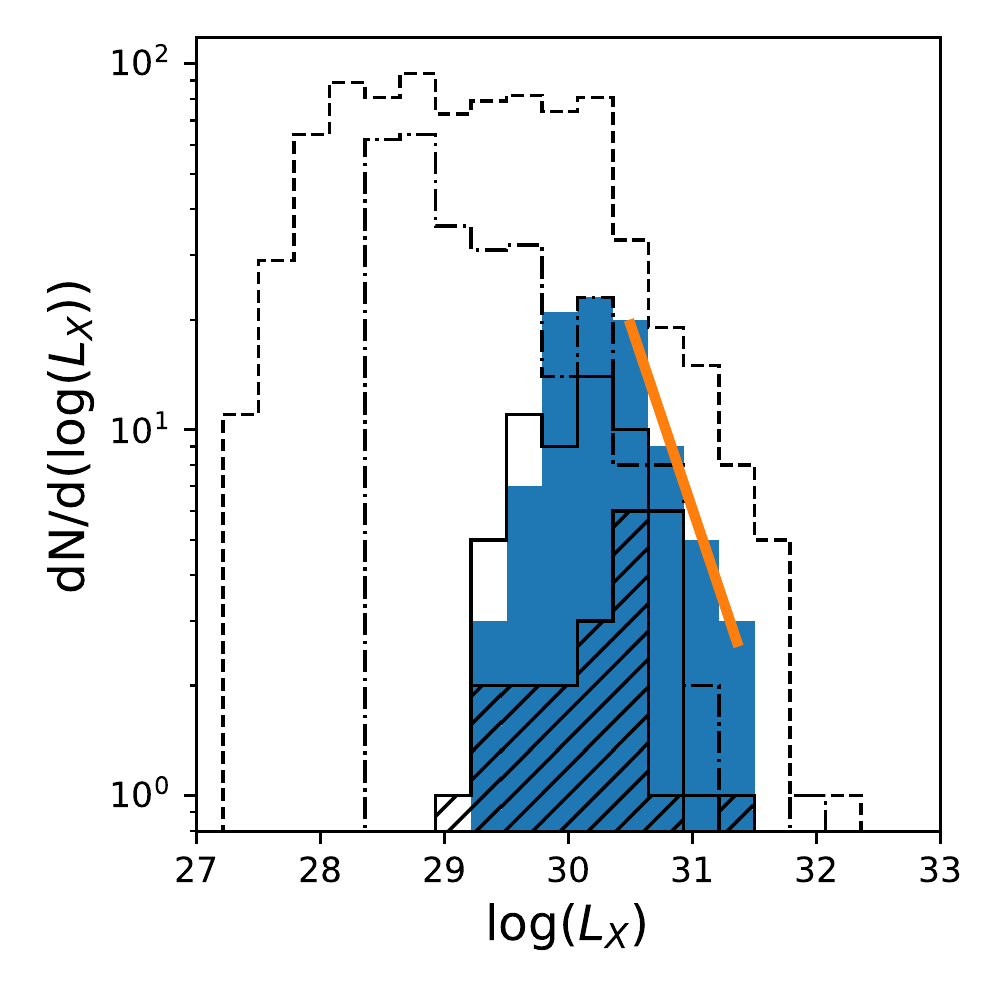}
    \caption{X-ray luminosity function of absorbed fluxes from X-ray-detected cluster members in IC 1396-West (solid line), in comparison to absorption-corrected fluxes from the adjacent IC 1396-A field (blue). Hatched bars represent sources with mid-IR data in our sample. The orange line fits the approximate XLF of \citet{2012MNRAS.426.2917G}. The \textit{XMM-Newton} observations do not detect cluster members as faint as those in \citet{2012MNRAS.426.2917G}. The XLF of 839 ONC stars \citep[dashed lines;][ the ``unobscured cool'' sample]{2005ApJS..160..379F} and IC 348 \citep[dot-dashed lines;][]{2012A&A...537A.135S} are shown for comparison.
    \label{fig:XLF}}
\end{figure}

To determine the X-ray sensitivity of the \textit{XMM-Newton} observation presented here, we converted the observed X-ray fluxes to luminosities using distances derived from \textit{Gaia} parallaxes. Only fifteen of our identified X-ray-detected cluster members have measured parallaxes, including thirteen with mid-IR data; for the other 43 sources, we make the assumption that they are at the estimated cluster distance of 931 pc. The X-ray source with the fewest counts detected of the X-ray-detected cluster members (i.e. sources with both near-IR and X-ray detections and are not outside the cluster per \textit{Gaia}) has X-ray flux $\sim 1.3 \times 10^{-15} \mathrm{erg}\,\mathrm{s}^{-1}\,\mathrm{cm}^{-2}$ and distance $914_{-159}^{+329}$ pc, corresponding to an absorbed $\log L_{\mathrm{X}} \approx 29.11$. For comparison, the faintest source with a determined $L_{\mathrm{X}}$ in \citet{2012MNRAS.426.2917G} has an unabsorbed $\log L_{\mathrm{X}} = 28.99 \pm 0.17$ and an $\log(N_{H}) = 20.26 \pm 0.32$, which yields an absorbed $\log L_{\mathrm{X}} \approx 28.95$. This indicates that the \textit{XMM} observations are less sensitive than the adjacent \textit{Chandra} field by $0.16$ dex. We present our absorbed XLF for both sources with mid-IR detections and the full set in comparison to the \citet{2012MNRAS.426.2917G} XLF in Figure \ref{fig:XLF}. We also compare these XLFs against those of the Orion Nebula Cluster \citep[the ``cool unobscured'' sample of cool stars with minimal optical extinction and X-ray absorption from][]{2005ApJS..160..379F} and IC 348 \citep{2012A&A...537A.135S}. 

Based on the comparison of our XLF to that of \citet{2012MNRAS.426.2917G}, we estimate that the IC 1396-West sample is complete only to an unabsorbed $\log L_{\mathrm{X}} \approx 30.8$, compared to the $\log L_{\mathrm{X}} \approx 30.5$ completeness level in the IC-1396A field. Using the $L_{X}-M$ relation of \citet{2007A&A...468..425T}, this yields a minimum stellar mass of $1.8 M_{\odot}$, well above the $1.0 M_{\odot}$ of \citet{2012MNRAS.426.2917G} and thus explaining the qualitative discrepancy between X-ray-detected disk fractions for the two data sets; indeed, \citet{2015A&A...576A..52R} note a clear difference in the rate of protoplanetary disk evolution (and thus disk lifetime) for $>2 M_{\odot}$ stars compared to $<2 M_{\odot}$ stars. We also note the similarity of the position of the XLF for both this observation of IC 1396 and that of \citet{2012MNRAS.426.2917G} to the XLF for IC 348, albeit not as complete at low $L_{X}$. We note that based on the exposure time for this \textit{XMM-Newton} observation, to achieve completeness down to $\log L_{\mathrm{X}} \approx 30.5$ would require a 174.5 ks observation, twice the length of the observation presented here. %We plot our X-ray luminosity function in comparison to that from \citet{2012MNRAS.426.2917G} in Figure \ref{fig:XLF}
%<insert discussion of XLF here>

\section{Spatial Structure of the IC 1396 Disk Distribution} \label{sec:others}

\begin{deluxetable*}{lccccccc}
\tablecaption{Distance Assessments for Sources Around IC 1396A From \citet{2012MNRAS.426.2917G} With \textit{Gaia} EDR3 \label{table:gaia_bins_getman}}
\tablehead{\colhead{Category} & \colhead{Num. Sources} & \colhead{Foreground} & \colhead{In Range} & \colhead{Background} & \colhead{Uncertain} & \colhead{No Plx} & \colhead{No Match}}
\startdata
DSK  &  51 &  2 & 26 &  1 &  9 &  9 &   4 \\
EXG  &  29 &  0 &  0 &  0 &  2 & 21 &   6 \\
FRG  &  21 & 10 &  4 &  1 &  4 &  1 &   1 \\
NOD  & 124 &  6 & 69 &  4 & 33 &  8 &   4 \\
UNC1 & 162 &  0 &  1 &  0 &  3 & 31 & 127 \\
UNC2 &  28 &  2 &  3 &  0 &  7 &  7 &   9 \\
\enddata
\tablecomments{Category definitions: DSK: disk-hosting source; EXG: extragalactic source; FRG: foreground; NOD: non-disk-host cluster member; UNC1: uncertain origin, likely extragalactic; UNC2: uncertain origin, possible YSOs.}
\end{deluxetable*}

The field observed with UKIRT/WFCAM fully overlaps with  \textit{Chandra}/ACIS-I observations of the cometary globule IC 1396A, presented in \citet{2012MNRAS.426.2917G}. To foster a direct comparison of our methods to that paper, we applied our methodology to the IC 1396-A field they observed, using their identifications of X-ray sources and the UKIRT/WFCAM data set to identify those objects that are likely cluster members. We note that as with the \textit{XMM-Newton} data set, we use a $2''$ search radius for matching all sources from \textit{Chandra} to the UKIRT/WFCAM dataset, while \citet{2012MNRAS.426.2917G} used a $2''$ match radius in the center and a $3.5''$ match radius further off-axis to match their X-ray sources with 2MASS. In 27 instances, \citet{2012MNRAS.426.2917G} identified a 2MASS match to an X-ray source that did not have a UKIRT counterpart within $2''$; these were cases where the X-ray source was off axis and the 2MASS match was between $2''$ and $3.5''$ away from the X-ray source. For these sources we searched for and identified a UKIRT counterpart to the 2MASS source, to ensure that we were evaluating approximately the same sources.

\citet{2012MNRAS.426.2917G} claimed 51 X-ray-detected disk hosts, and 124 X-ray-detected non-disk-host cluster members, resulting in a disk fraction of $29^{+4}_{-3}\%$. By contrast, applying our methodology to the same field results in 38 X-ray-detected disk hosts and 75 X-ray-detected non-disk-host cluster members, resulting in a disk fraction of $33^{+5}_{-4}\%$.

We hypothesize that the difference in our estimated numbers of disk hosts and non-disk-host cluster members is due to our use of UKIRT/WFCAM data and \textit{Gaia} EDR3 parallaxes, while \citet{2012MNRAS.426.2917G} used 2MASS data and estimated cluster membership based on color cuts and the X-ray data, since \textit{Gaia} data was not available at their time of publication to provide an independent distance constraint. To check the effect of \textit{Gaia} parallaxes, we independently estimated distances to their catalog of sources using \textit{Gaia} EDR3 parallaxes. \citet{2012MNRAS.426.2917G} broke the X-ray source classifications in the \textit{Chandra} field into six categories: DSK (a disk-hosting cluster member), EXG (extra-galactic source), FRG (a foreground source), NOD (a non-disk-hosting cluster member), and UNC1 and UNC2 (two different types of uncertain member). Based on the 90\% confidence intervals in distance for each object (using the same methodology for determining distances and distance uncertainties as Section \ref{sec:classifications} compared to the distance range derived in Section \ref{sec:gaiaedr3}) we sorted objects in the \textit{Chandra} field into six bins: \textit{in-range} (distance interval overlaps with the cluster distance); \textit{foreground} (no overlap; entire confidence interval is in foreground); \textit{background} (no overlap; entire confidence interval is in background); \textit{uncertain} (overlaps, but with distance uncertainty ranges $>1000$ pc), \textit{no parallax} (no parallax measured by \textit{Gaia}), and 151 sources with \textit{no match} between \textit{Chandra} and UKIRT; while the majority of these come from the 219 sources in categories that had faint or no IR counterparts (EXG, UNC1, UNC2), typically expected to be extragalactic \citep[see Section 3.2 of][]{2012MNRAS.426.2917G}, some of these are also sources undetected in 2MASS/UKIRT but detected with \textit{Spitzer}/IRAC. We present these results in Table \ref{table:gaia_bins_getman}, but focus our attention on the 121 sources in the ``FRG,'' ``DSK,'' and ``NOD'' classifications with UKIRT matches and \textit{Gaia} parallaxes that yield a distance with uncertainty $<1000$ pc, as these are the most certain classifications.

We find that 90\% of the DSK sources with UKIRT matches and \textit{Gaia} EDR3-based distance uncertainties $<1000$ pc have distances consistent with the cluster distance. Similarly, 87\% of the NOD sources with UKIRT matches have \textit{Gaia} parallaxes consistent with the cluster distance. Only 67\% of FRG sources are actually in the foreground, while four are at the correct distance for the cluster. We also find that our YSO-class assignments disagree in a handful of cases with those of \citet{2012MNRAS.426.2917G}---we classify three of their disk-hosts as Class III YSOs, and seven of their Class III YSOs as disk hosts. The remaining disks identified by our search that were not identified in \citet{2012MNRAS.426.2917G} emerge from the UNC1 and UNC2 bins, where the presence of UKIRT data and additional information from \textit{Gaia} proves decisive. On the whole, while \textit{Gaia} data does improve constraints on the distances, the classifications of \citet{2012MNRAS.426.2917G} are reasonably accurate, demonstrating the utility of this method of cluster member identification for clusters at distances where \textit{Gaia} data is incomplete.

While the number of sources in IC 1396-West is small enough that a robust statistical comparison of the \textit{XMM-Newton} field with the \textit{Chandra}/ACIS-I field is infeasible, we can qualitatively compare these two regions, as well as the cluster core region analyzed by \citet{2009AJ....138....7M} (which we do not re-analyze here due to that field falling outside the UKIRT field). The highest estimated fraction of X-ray-detected YSOs that have disks is that of \citet{2012MNRAS.426.2917G}, which might suggest an increase in disk fraction from the cluster core to this field. However, we think it is more likely that this is because the observations analyzed by \citet{2009AJ....138....7M} were made with the HETG grating in place on \textit{Chandra}, which significantly diminishes the efficiency of the observations and likely led to non-detection of some sources. Overall, the data indicate that the disk fraction \textit{does not increase} as a function of distance from the cluster core. Future work with deeper X-ray observations will be necessary to fully quantify this effect in clusters other than the ONC.

%The uncertainties on these distances will be substantially lowered with \textit{Gaia} EDR3, which will affect these categorizations. However, as \citet{2019A&A...622A.118S} does not list which sources they used to determine the distance to IC 1396A, we cannot easily compare the distance range listed there to the \textit{Gaia} EDR3 data, as the distance range estimate to IC 1396A apparently does not correct for the \textit{Gaia} DR2 zero-point \citep{2018A&A...616A...2L}, as was done in \citet{2018AJ....156...58B}.

%% can we use some  bullet points for summary?
%% Done!
\section{Summary} \label{sec:conclusion}

In this paper, we present X-ray detections from an 87.3-ks exposure with the \textit{XMM-Newton} Observatory, which we combine with near-infrared wide-field images with UKIRT/WFCAM, and archival data from \textit{Gaia} EDR3, \textit{Spitzer}, and AllWISE to identify disk-hosting and non-disk-hosting young stars forming in a field at the western edge of the Elephant's Trunk Nebula, which we designate as IC 1396-West. We also used \textit{Gaia} EDR3 to re-evaluate the color-cut-based YSO identification scheme of \citet{2012MNRAS.426.2917G} for \textit{Chandra} observations of the field surrounding the cometary globule IC 1396A. We find the following:

\begin{itemize}
    \item We identify 421 new members of the Tr 37/IC 1396A cluster, including 86 Class I YSOs (one detected in X-rays), 51 Class FS YSOs (one detected in X-rays), 229 Class II YSOs (one detected in X-rays), 15 Class III YSOs (all detected in X-rays), and 40 sources with X-ray detections but no mid-IR data to determine a YSO class.

    \item The disk fraction of X-ray-detected YSOs in IC 1396-West is $17_{-7}^{+10}\%$,  statistically consistent with the $29_{-3}^{+4}\%$ rate for an adjacent region closer to the core of IC1396, centered on the globular cluster IC1396A and observed with \textit{Chandra} ACIS-I \citep{2012MNRAS.426.2917G}. The \textit{XMM-Newton} X-ray observations do not probe to as deep of a luminosity as the \textit{Chandra} observations of IC 1396A, which leads to those observations extending to lower completeness than the \textit{XMM-Newton} observations here. The small number of sources detected, due in part to both the difference in completeness of the observations and the lower sensitivity of the mid-IR data for most of IC 1396-West (mostly \textit{WISE}) compared to the IC 1396A field (fully observed by \textit{Spitzer}) makes any differences statistically insignificant. This result also does not consider lower-mass stars too faint to be detected in the \textit{XMM-Newton} or \textit{Chandra}/ACIS observations that retain their disk longer than the higher-mass stars we detect in X-rays.

    \item We note a surprising number of Class I YSOs along the nebula, suggesting ongoing star formation in this region later than the birth of stars in Tr 37.

    \item While color-cut-based searches for YSOs in clusters without distance information is reasonably effective, the addition of independent distance measurements provides key additional constraints on YSO identification. We find from our re-analysis of the field of \textit{Chandra} observations of IC 1396A that the disk fraction following our methodology is $33^{+5}_{-4}\%$, comparable with the $29_{-3}^{+4}\%$ rate derived by \citet{2012MNRAS.426.2917G}.
\end{itemize}%at  We find that   , finding that 

While the additional information provided by \textit{Gaia} was invaluable to us in our search for Class III YSOs, it was not entirely conclusive, as \textit{Gaia} parallaxes are not complete to the distance of Tr 37, especially with particularly dense background emission from the nebula. We expect that future releases from \textit{Gaia} will yield parallax information for more stars, which will provide better identification of Class III sources---in particular, those sources with photospheric SED slopes and detected X-ray emissions but no parallax measurements. This improved completeness will also enable more effective determination of what fraction of sources with photospheric SEDs at the correct distance but are not detected in X-rays are likely to be members of the cluster, which will further improve estimates of the age of clusters by extending membership lists to lower luminosities.

Even after the end of its mission, the \textit{Gaia} data set will have limitations that can be overcome by the use of multiple observation methods. The X-ray observations we collected here show a set of potential YSOs (based on their position relative to the nebula and their X-ray hardness ratios) that went unidentified by \textit{Gaia}, demonstrating the utility of an approach that combines data from \textit{Gaia} and other observatories. As the large X-ray catalog from the eROSITA mission \citep{2014SPIE.9144E..1TP} becomes available, this data will be crucial for helping to identify the objects that are missed by \textit{Gaia}, helping us build a more comprehensive picture of the formation and evolution of young stellar objects.


\acknowledgements

We thank the anonymous reviewer for providing comments that helped to improve the content and clarity of this paper. The authors thank Aurora Sicilia-Aguilar for valuable discussion on estimating the distance to IC 1396. The authors thank Nicholas J.G.\ Cross for helpful comments on the WFCAM archive. SMS and HMG were supported by the National Aeronautics and Space Administration through Chandra Award Number GO9-20019X issued by the Chandra X-ray Observatory Center, which is operated by the Smithsonian Astrophysical Observatory for and on behalf of the National Aeronautics Space Administration under contract NAS8-03060. SJW was supported by NASA contract NAS8-03060.

When the data reported here were acquired, UKIRT was supported by NASA and operated under an agree-ment  among  the  University of  Hawaii,  the  University of Arizona, and Lockheed Martin Advanced Technology Center;  operations  were  enabled  through the  cooperation of the East Asian Observatory. UKIDSS uses the UKIRT Wide Field Camera \citep[WFCAM;][]{2007A&A...467..777C} and a photometric system described in \citet{2006MNRAS.367..454H}. The pipeline processing and science archive are described in \citet{2008MNRAS.384..637H}.

This work is based in part on observations made with the Spitzer Space Telescope, which is operated by the Jet Propulsion Laboratory, California Institute of Technology under a contract with NASA.

This publication makes use of data products from the Wide-field Infrared Survey Explorer, which is a joint project of the University of California, Los Angeles, and the Jet Propulsion Laboratory/California Institute of Technology, funded by the National Aeronautics and Space Administration.

This research has made use of the NASA/IPAC Infrared Science Archive, which is funded by the National Aeronautics and Space Administration and operated by the California Institute of Technology.

This work has made use of data from the European Space Agency (ESA) mission
{\it Gaia} (\url{https://www.cosmos.esa.int/gaia}), processed by the {\it Gaia}
Data Processing and Analysis Consortium (DPAC,
\url{https://www.cosmos.esa.int/web/gaia/dpac/consortium}). Funding for the DPAC
has been provided by national institutions, in particular the institutions
participating in the {\it Gaia} Multilateral Agreement.

This research has made use of the SIMBAD database, operated at CDS, Strasbourg, France. This research has made use of the VizieR catalogue access tool, CDS, Strasbourg, France. This research made use of the cross-match service provided by CDS, Strasbourg.

This publication makes use of data products from the Two Micron All Sky Survey, which is a joint project of the University of Massachusetts and the Infrared Processing and Analysis Center/California Institute of Technology, funded by the National Aeronautics and Space Administration and the National Science Foundation.

%The Pan-STARRS1 Surveys (PS1) and the PS1 public science archive have been made possible through contributions by the Institute for Astronomy, the University of Hawaii, the Pan-STARRS Project Office, the Max-Planck Society and its participating institutes, the Max Planck Institute for Astronomy, Heidelberg and the Max Planck Institute for Extraterrestrial Physics, Garching, The Johns Hopkins University, Durham University, the University of Edinburgh, the Queen's University Belfast, the Harvard-Smithsonian Center for Astrophysics, the Las Cumbres Observatory Global Telescope Network Incorporated, the National Central University of Taiwan, the Space Telescope Science Institute, the National Aeronautics and Space Administration under Grant No. NNX08AR22G issued through the Planetary Science Division of the NASA Science Mission Directorate, the National Science Foundation Grant No. AST-1238877, the University of Maryland, Eotvos Lorand University (ELTE), the Los Alamos National Laboratory, and the Gordon and Betty Moore Foundation.

\vspace{5mm}
\facilities{IRSA, Spitzer, WISE, UKIRT/WFCAM, XMM-Newton}


\software{AstroPy \citep{2013A&A...558A..33A,2018AJ....156..123A}, NumPy \citep{van2011numpy,harris2020array}, SciPy \citep{jones_scipy_2001,2020SciPy-NMeth}, Matplotlib \citep{Hunter:2007}, pandas \citep{mckinney,jeff_reback_2020_3644238}}



% \appendix


\bibliography{references}{}
\bibliographystyle{aasjournal}



\end{document}